\documentclass{article}
\usepackage{spconf,amsmath,graphicx}
\usepackage{graphicx}
\usepackage{multirow}
\usepackage{booktabs}
\usepackage{tabularx}
\usepackage{subfigure}
\usepackage{hyperref}

\usepackage{amsmath,amsfonts,bm}

\def\eqref#1{equation~\ref{#1}}
\def\1{\bm{1}}

\def\vw{{\bm{w}}}

\DeclareMathAlphabet{\mathsfit}{\encodingdefault}{\sfdefault}{m}{sl}
\SetMathAlphabet{\mathsfit}{bold}{\encodingdefault}{\sfdefault}{bx}{n}

\title{Differentiable Wavetable Synthesis}
\name{Siyuan Shan$^{1, \star}$ \thanks{$^{\star}$ Work performed while a PhD intern at ByteDance. Corresponding Email: \textit{siyuanshan@cs.unc.edu}} \qquad Lamtharn Hantrakul$^{2}$ \qquad Jitong Chen$^{2}$\qquad Matt Avent$^{2}$\qquad David Trevelyan$^{2}$}
  
  \address{$^{1}$ University of North Carolina at Chapel Hill \\
      $^{2}$ Speech, Audio \& Music Intelligence (SAMI) Team, ByteDance}

\begin{document}
%\ninept
%
\maketitle
\begin{abstract}
Differentiable Wavetable Synthesis (DWTS) is a technique for neural audio synthesis which learns a dictionary of one-period waveforms i.e. wavetables, through end-to-end training. We achieve high-fidelity audio synthesis with as little as 10 to 20 wavetables and demonstrate how a data-driven dictionary of waveforms opens up unprecedented one-shot learning paradigms on short audio clips. Notably, we show audio manipulations, such as high quality pitch-shifting, using only a few seconds of input audio. Lastly, we investigate performance gains from using learned wavetables for realtime and interactive audio synthesis. 
\end{abstract}
%
%%% LEARNED WAVETABLES FIGURE %%%
\begin{figure*}[t!]
\centering
\includegraphics[width=0.7\textwidth]{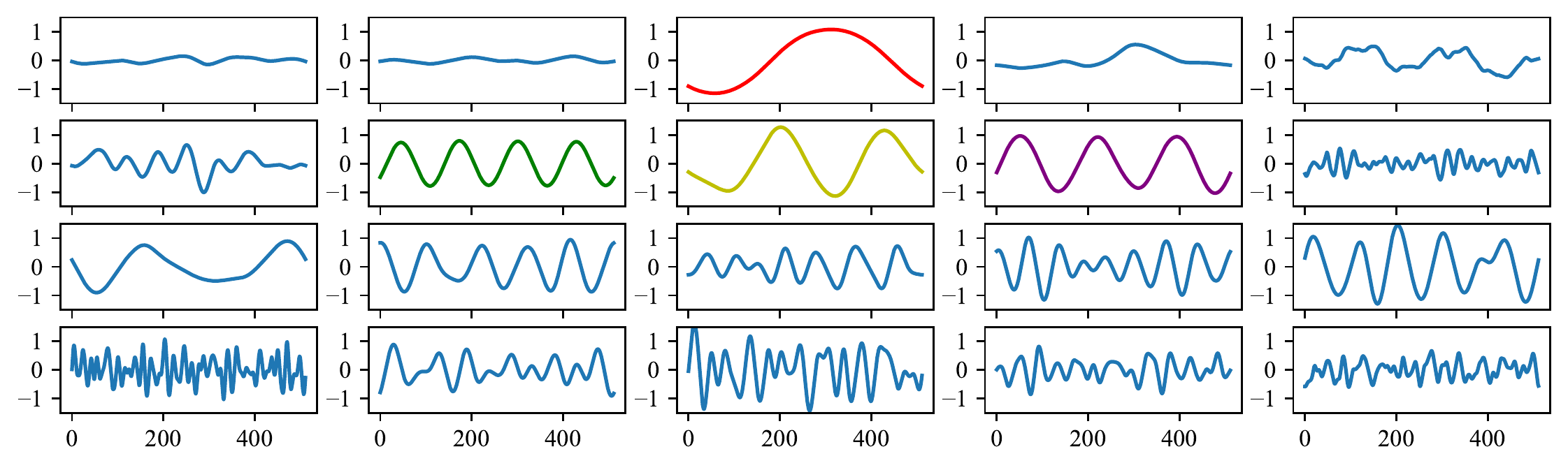}
\caption{Learned wavetables ordered with highest average attention weights appearing first (normal English reading order). Wavetables of key harmonics are highlighted:  $f_0$ (red), $f_1$ (yellow), $f_2$ (purple) and $f_3$ (green). The remaining wavetables are data-driven combinations of higher harmonics. The first two wavetables appear to be silence.}
\label{fig:learned_wavetables}
\end{figure*}

\section{Introduction}
\label{sec:Introduction}

Although machine learning (ML) has revolutionized modern audio synthesis with unprecedented progress \cite{oord2016wavenet, wang2017tacotron, Kalchbrenner2018wavernn}; fast, robust and realtime neural audio synthesis remains a challenge. Purely deep audio models require significant parallelization \cite{ren2021fastspeech2} or custom kernels \cite{Kalchbrenner2018wavernn} for fast performance. Recent techniques fusing differentiable signal processing with deep neural networks (DNN) \cite{engel2020ddsp} have enabled efficient, interpretable and real-time use cases \cite{ganis2021realtime-ddsp, kuznetsov2020diffiir} with minimal additional engineering.

Wavetable synthesis (WTS) is well-suited to realtime synthesis of periodic and quasi-periodic signals. Real-world objects that generate sound often exhibit physics that are well described by harmonic oscillations. These include vibrating strings, membranes, hollow pipes and human vocal chords \cite{smith2006physical-audio-signal-processing}. By using lookup tables composed of single-period waveforms, WTS can be as general as additive synthesis whilst requiring less realtime computation \cite{bristow1996wavetable}. We demonstrate three key contributions to neural audio synthesis using this approach: high fidelity synthesis, wavetables transferrable to other tasks and an order of magnitude reduction in computational complexity. 

\section{Related Work}
\label{sec:Related Work}
\textbf{Wavetable Synthesis (WTS):} Wavetable synthesis generates audio from a collection of static, single-period waveforms called ``wavetables", that each capture a unique harmonic spectrum. Wavetables are typically 256 - 4096 samples in length; a collection can contain a few to several hundred wavetables depending on use case. Periodic waveforms are synthesized by indexing into the wavetables as a lookup table and interpolating between neighbouring samples. WTS was historically used on early hardware synthesizers where memory and computing power were limited. Today, WTS continues to underpin commercial sound design and synthesis tools due to its ability to generate a wide variety of timbres coupled with efficient performance. Wavetables are normally hand-crafted or extracted programatically from audio spectra \cite{bristow1996wavetable, Horner2008}. In this work, we learn data-driven wavetables.

\textbf{Differentiable dictionaries:} ML models can be equipped with differentiable memory or dictionaries that are learned end-to-end with model parameters. Models can write to external memory to record and lookup changes in state \cite{graves2014neural}. Discriminative models can learn a differentiable dictionary for improved classification during inference \cite{shan2020meta}. Here, we formulate wavetables as a differentiable dictionary optimized jointly with model parameters. 

\textbf{Differentiable Digital Signal Processing:} DDSP \cite{engel2020ddsp} describes a family of techniques utilizing strong inductive biases from DSP combined with modern ML. Successful applications include audio synthesis \cite{engel2020ddsp}, pitch detection \cite{engel2020inv-ddsp-pitch}, artificial reverberations \cite{lee2021diffreverb}, IIR filters \cite{kuznetsov2020diffiir} and audio effect manipualtion \cite{ramirez2021differentiable}. Building on these works, we add WTS as a new technique for generative audio tasks.

\section{Methods}
\label{sec:Methods}

\subsection{Differentiable Wavetable Synthesizer (DWTS)} 
\textbf{Wavetable as learnable dictionary:} We define a learnable dictionary $D=\{\vw_{i}\}_{i}^{N}$ where $N$ is the number of wavetables and $\vw_{i} \in \mathbb{R}^L$ denotes a one-cycle wavetable of length $L$. When a wavetable begins and ends on different values, this discontinuity causes synthesis artefacts. We append $\vw_i[L+1]$ to $\vw_i$ and set $\vw_i[L+1] = \vw_i[0]$. A wavetable $\vw_i$ now contains $L+1$ elements with $L$ learnable parameters. During training, we learn $D$ using gradient descent jointly with other parameters. During inference, $D$ is frozen and treated as a traditional, static collection of wavetables. 

\textbf{Time-varying attention:} By sequentially morphing between wavetables, timbre can be changed over time \cite{Horner2008}. Inspired by \cite{horner1995genetic-algorithm-wavetable} we generalize morphing as a time-varying linear attention over all wavetables i.e. $\ c_{1}^{N}, c_{2}^{N} ... c_{T}^{N}$ where $N$ and $T$ are number of wavetables and timesteps respectively with constraints $\sum_{i=1}^{N}c_i(n)=1$ and $c_i(n)\geq 0$.

\textbf{Phase:} Our model ``draws" wavetables directly in the time domain. Phase relationships within and across wavetables can be controlled without needing to coherently manage independent magnitudes and phases in the complex frequency domain. This contrasts with \cite{engel2020ddsp}, where the initial phase of all harmonic components are fixed at 0.

\textbf{Synthesis}: At the heart of WTS is a phase accumulator \cite{bristow1996wavetable}. Given an input sequence of time-varying fundamental frequency $f_0(n)$ over discrete time steps $n$, we can compute the instantaneous modulo phase $\tilde{\phi}$ by integrating $f_0(n)$:

\begin{equation}
\label{eq:phase}
    \tilde{\phi}(n) = 2 \pi \sum_{m=0}^{n}f_0(m) \mod 2\pi .
\end{equation}

$\tilde{\phi}(n)$ is normalized into a fractional index $\tilde{j}(n) = \frac{L}{2\pi}\tilde{\phi}(n)$. We synthesize the signal $x(n)$ by linearly combining wavetables $\vw_i$ in $D$ via:
\begin{equation}
\label{eq:synthesize}
    x(n) = A(n) \sum_{i=1}^{N}c_i(n)\cdot\Phi(\vw_i, \tilde{j}(n), \kappa),
\end{equation}

where $A(n)$ is a time-varying amplitude controlling the signal's overall amplitude and $c_i$ denotes the time-varying attention on $\vw_i$. $A(n)$ and $c_i(n)$ are constrained positive via a sigmoid. The function $\Phi(\vw_i, \tilde{j}, \kappa)$ is a fractional indexing operator that returns the $(\tilde{j})$-th element of the vector $\vw_i$ by using an interpolation kernel $\kappa$ to approximate $\vw_i[\tilde{j}]$ when $\tilde{j}$ is non integer. Whilst more sophisticated interpolation kernels exist (cubic, spline etc.), we use linear interpolation for optimal realtime performance. For readers unfamiliar with fractional indexing, please refer to the online supplement. 

\textbf{Initialization:} 
$\vw_i$ is randomly initialized with a zero-centered Gaussian distribution $\mathcal{N}(0, \sigma^2)$. We empirically find using a small $\sigma=0.01$ improves training.

\textbf{Antialiasing:} At high $f_0$, the frequencies of upper harmonics in a wavetable can be above Nyquist and must be removed before lookup to prevent aliasing. This filter also prevents high frequency components present in the Gaussian noise initialization from causing aliasing at the start of training. Without this frequency-dependent anti-aliasing filter, we found aliasing artefacts alone prevented meaningful learning.

\begin{table*}
\centering

\caption{Reconstruction error comparison on the Nsynth dataset. Compared to the SOTA additive synthesis approach \cite{engel2020ddsp}, our wavetable synthesis approach achieves comparable or lower errors.}
\label{table:1}
\setlength{\tabcolsep}{2.5pt}
\begin{tabular}{@{}c|cccc@{}}

\toprule
       \multirow{2}{*}{DDSP Additive Synthesis} & \multicolumn{4}{c}{Wavetable Synthesis (Ours) with $N$=}                                                                                                              \\
                                                         & 5        & 10            & 20            & 100                  \\ \midrule

                                     0.5834 $\pm$ 0.0035        & 0.6448  $\pm$ 0.0041          & 0.5989 $\pm$ 0.0042            & \textbf{0.5712 $\pm$ 0.0037}         & 0.5756 $\pm$ 0.0034              \\
   
                              \bottomrule
\end{tabular}
\end{table*}

%%% RECONSTRUCTION SPECTROGRAM FIGURE%%%
\begin{figure}[t!]
\subfigure[Target]{\includegraphics[width=0.24\linewidth, trim={1.45cm 0.2cm 1cm 0.2cm},clip]{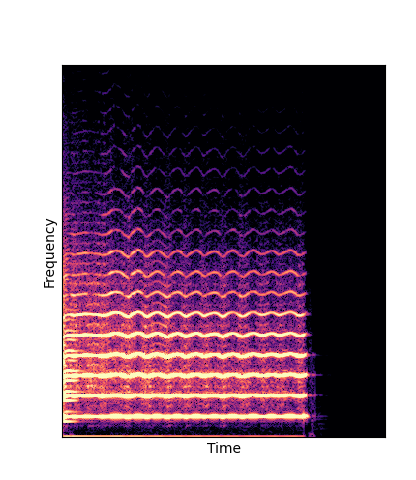}}
\subfigure[Reconstructed]{\includegraphics[width=0.24\linewidth, trim={1.45cm 0.2cm 1cm 0.2cm},clip]{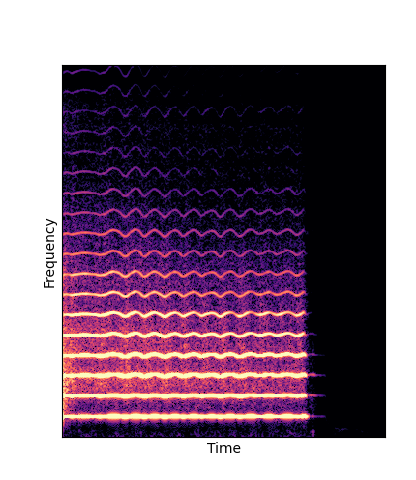}}
\subfigure[Target]{\includegraphics[width=0.24\linewidth, trim={1.45cm 0.2cm 1cm 0.2cm},clip]{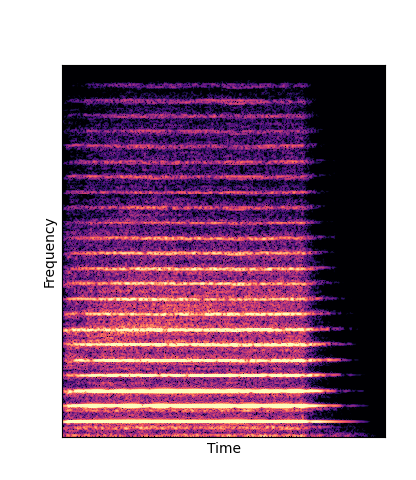}}
\subfigure[Reconstructed]{\includegraphics[width=0.24\linewidth, trim={1.45cm 0.2cm 1cm 0.2cm},clip]{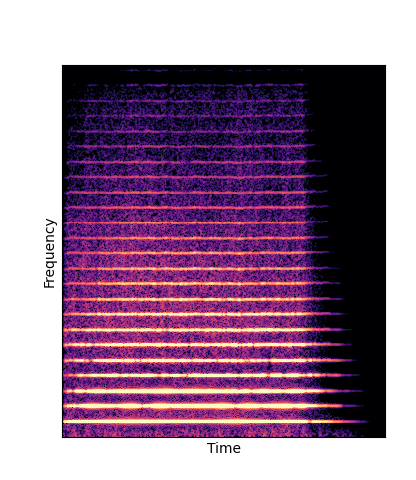}}

%\vspace{-10pt}
  \caption{Spectrograms of two target samples and their corresponding reconstruction.}
  \label{fig:recon}
\end{figure}

\section{Experiments}

We benchmark against the original DDSP autoencoder \cite{engel2020ddsp}, where a DNN controls an additive synth and filtered noise synth to produce harmonic and non-harmonic components of audio respectively. We replace the additive synth with our wavetable synth and use an identical filtered noise synth. Noise generation is a stochastic process that must be modelled separately. Like \cite{engel2020ddsp}, we omit the optional reverb module when training on the NSynth dataset \cite{engel2017neural}.

\textbf{Model:} We adopt an identical input tuple ($f_0(n), l(n), z(n)$) \cite{engel2020ddsp}. Fundamental frequency $f_0(n)$ in \eqref{eq:phase} is extracted by a pretrained CREPE model \cite{kim2018crepe} with fixed weights. Loudness $l(n)$ is an A-weighted log-magnitude extracted deterministically from audio. The residual embedding $z(n)$ is extracted from MFCC's via an encoder. For direct comparison,  we use identical encoder and decoder architectures with approximately 7M parameters in total.

Unlike the autoencoder in \cite{engel2020ddsp}, our setup contains an additional $N\times L$ learnable parameters in the wavetable dictionary $D$ during training. During inference however, the wavetables are frozen and the parameter counts near-identical.

\textbf{Loss:} We use a multi-scale spectral loss similar to \cite{engel2020ddsp}:
\begin{equation}
    L_{\text{reconstruction}} = \sum_{i} ||S_i - \hat{S}_i ||_1,
\end{equation}
where $S_i$ and $\hat{S}_i$ respectively denote magnitude spectrums of target and synthesized audio, and $i$ denotes different FFT sizes. We found the log term $||\log S_i - \log \hat{S}_i ||_1$ caused training instabilities and excluded it. This modification did not influence the quality of synthesized audio.

\textbf{Dataset} We use the same subset of the NSynth dataset \cite{engel2017neural} in \cite{engel2020ddsp,  hantrakul2019fast, engel2019gansynth}, containing ~70,000 mono 16kHz samples each 4 seconds long. The examples comprise mostly of strings, brass, woodwinds and mallets. At 16kHz, a wavetable length $L = 512$ is enough to represent all harmonics at the lowest fundamental frequency of interest (20Hz).

\section{Results}
\label{sec:Results}

\subsection{Reconstruction Quality of DWTS}
Table \ref{table:1} reports the reconstruction error of the SOTA additive-based autoencoder from \cite{engel2020ddsp} and the proposed DWTS-based autoencoder. We vary the number of wavetables $N=5,10,20,100$. Our approach achieves the lowest reconstruction error of 0.5712 using only 20 wavetables. At the expense of a small reduction in quality compared to the baseline, $N$ can be low as 10. Fig \ref{fig:recon} shows spectrograms of two samples and their matching reconstructions using wavetables. We encourage readers to listen to the online supplement\footnote{https://lamtharnhantrakul.github.io/diffwts.github.io/}.

Crucially, the wavetables in $D$ form an alternative, compact set of basis vectors spanning an $L$-dimensional space extracted directly from the data. When $N$ is very small at 5, reconstruction suffers due to an insufficient number of bases. 10-20 wavetables strike an optimal balance for the NSynth dataset. Wavetables reduce the number of control dimensions by an order of magnitude compared to the 100 sinusoids in an additive synth \cite{engel2020ddsp}. More importantly, the extracted wavetables are an explicit dictionary that are portable to other tasks. We show this property in later sections.

\subsection{Visualizing Wavetables} 
\begin{figure}[t!]
\centering
\includegraphics[width=0.4\textwidth]{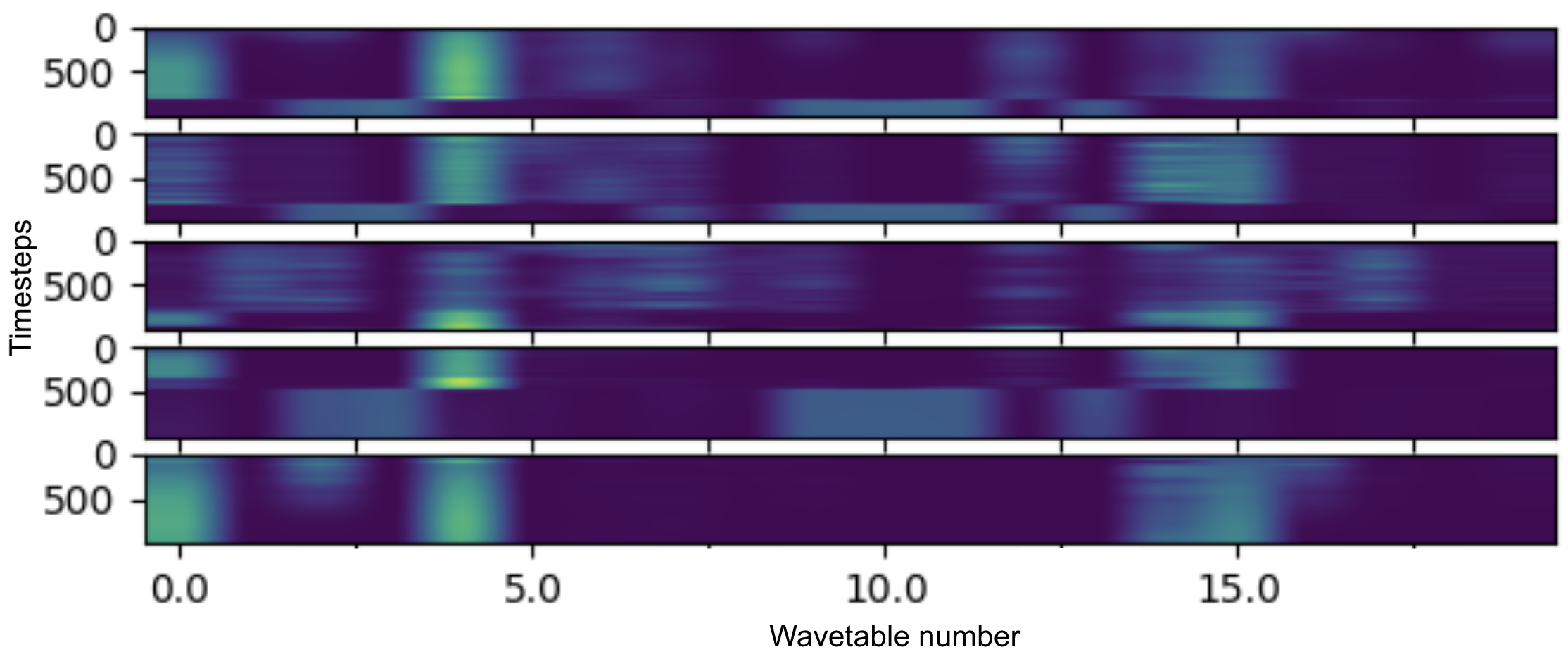}
\caption{Visualization of the time-varying attention weights of 5 samples using 20 wavetables.}
\label{fig:attention}
\end{figure}

Fig \ref{fig:learned_wavetables} shows learned wavetables from the NSynth dataset when $N=20$. Despite being initialized with noise, the learned wavetables are smooth and diverse in shape. They also match the physics of NSynth sounds. In acoustic instruments, energy is focused on lower frequencies, particularly the first few harmonics, compared to higher harmonics \cite{Rossing1998PhysicsMusicalInstruments}. Wavetables in Fig \ref{fig:learned_wavetables} are ordered with highest average attention weights appearing first. The wavetable highlighted in red is a phase-shifted sinusoid of one period i.e. the fundamental frequency $f_0$. Other key partials  $f_1$, $f_2$ and $f_3$ are highlighted in yellow, purple and green. The remaining wavetables are data-driven combinations of higher harmonics, compactly summarizing in a single wavetable entry what would have taken several sinusoidal components in an additive synth \cite{engel2020ddsp} to represent. Fig \ref{fig:attention} shows the attention weights over time for five audio samples. The attention visibly shifts across wavetables to output the desired spectrum.

\begin{figure*}[t]
\center
\subfigure[Original]{\includegraphics[width=0.15\linewidth, trim={1.45cm 0.2cm 1cm 0.2cm},clip]{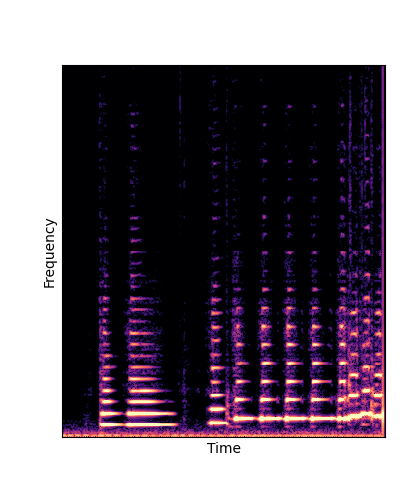}}
\subfigure[\textbf{Add  Scratch}]{\includegraphics[width=0.15\linewidth, trim={1.45cm 0.2cm 1cm 0.2cm},clip]{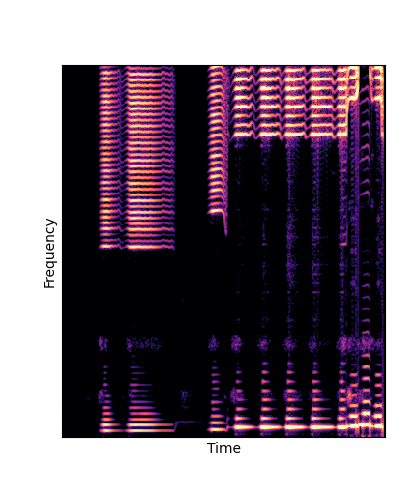}}
\subfigure[\textbf{Add Pretrain}]{\includegraphics[width=0.15\linewidth, trim={1.45cm 0.2cm 1cm 0.2cm},clip]{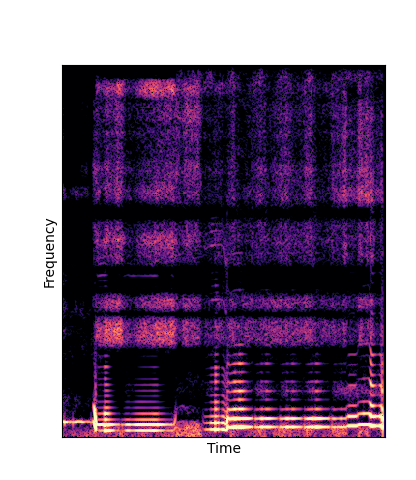}}
\subfigure[\textbf{DWTS Scratch}]{\includegraphics[width=0.15\linewidth, trim={1.45cm 0.2cm 1cm 0.2cm},clip]{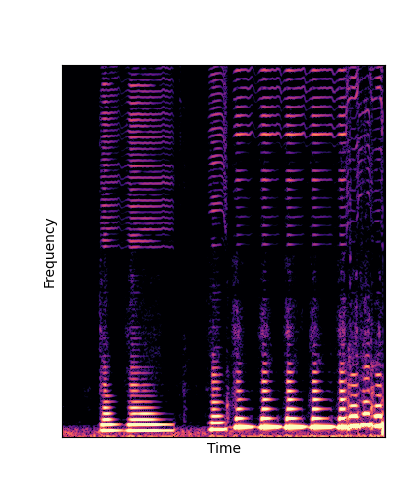}}
\subfigure[\textbf{DWTS Pretrain}]{\includegraphics[width=0.15\linewidth, trim={1.45cm 0.2cm 1cm 0.2cm},clip]{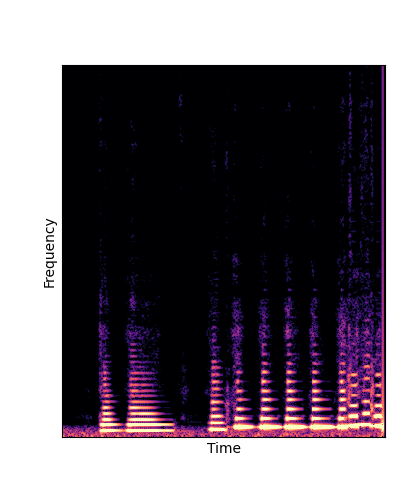}}

%\vspace{-10pt}
  \caption{Spectrograms of original audio (a) and synthesized samples from an input $f_0(n)$ pitch shifted down by an octave (b-e)}
  \label{fig:oneshot}
\end{figure*}

The asymmetric wavetables are the result of complex behavior in magnitude and phase. We found phase-locking wavetables to start and end at 0 deteriorated performance. It suggests the model takes advantage of phase relationships within and between wavetables. This precise control of wavetable phase will be particularly valuable in future work exploring synthesis of stereo and binaural audio \cite{richard2020binaural}.

\subsection{One shot learning and audio manipulations}  
In domains such as Neural Machine Translation (NMT), word embeddings \cite{mikolov2013word2vec} extracted from end-to-end language learning are useful for other language tasks. Analogously, we reason our data-driven wavetables should be useful in other scenarios like one-shot learning and data-efficient extrapolation. Unlike an implicit multi-dimensional vector, wavetables are an explicit and interpretable representation.

\textbf{One-shot setup:} Given only a single 4 second passage of saxophone from the URMP dataset \cite{li2018creating}, we train a new autoencoder model initialized with pretrained wavetables from Fig \ref{fig:learned_wavetables} (\textit{DWTS Pretrain}). This model only outputs time-varying attention weights, since the wavetables are now a fixed dictionary lookup. We compare against three baselines: (1) additive-synth autoencoder trained from scratch (\textit{Add Scratch}), (2) finetuning an additive-synth autoencoder pretrained on Nsynth (\textit{Add Pretrain}) and (3) Wavetable-synth autoencoder trained from scratch (\textit{DWTS Scratch}).

\textbf{Pitch extrapolation:} While all models achieve identical high quality one-shot reconstructions of the saxophone segment, only \textit{DWTS Pretrain} is robust to overfitting during extrapolation.  Fig \ref{fig:oneshot} shows how all baselines exhibit high frequency artefacts when input $f_0(n)$ is shifted. \textit{DWTS+Pretrain} remains artefact free even at extreme shifts, such as 3 octaves below the original sample. 

We repeat with a 4 second piano passage. A piano is challenging to model due to the presence of both many harmonics and percussive hammer hits \cite{Rossing1998PhysicsMusicalInstruments}. We also compare against the Librosa library pitch shift function based on traditional DSP \cite{mcfee2015librosa}. When resynthesizing the segment 3 octaves down, \textit{DWTS Pretrain} is the only method that preserves the hammer's percussive impact and independently shifts harmonic components. Librosa pitch shift loses the transient impact completely. 

\textbf{Connection to PSOLA:} In this scenario, we hypothesize \textit{DWTS Pretrain} approaches an optimal Pitch Synchronous Overlap and Add (PSOLA) algorithm \cite{charpentier1986psola}. PSOLA windows a single cycle of the original waveform in the time domain, re-patching and overlapping these windows at the new pitch. Imperfections in this windowing and overlapping process can cause artefacts. \textit{DWTS Pretrain} guarantees single-cycle waveforms in $D$, where re-pitching is trivially achieved using a slower phase accumulator $\tilde{\phi}(n)$ reading through a wavetable $\vw_i$. This opens up applications like data-efficient neural sampling, pitch correction and efficient polyphony using multiple phase accumulators. We leave wavetables extracted from speech or singing voice for future work.

\subsection{Computational Complexity}  
Realtime performance is determined by
many factors including algorithmic complexity, low level optimizations,
target hardware and the C++ ML framework used for matrix operations. Here, we consider only the key difference between
the additive synth-based pipeline in \cite{engel2020ddsp}, which already runs realtime \cite{ganis2021realtime-ddsp}, and our new
approach.

A.) \textit{Additive}: a bank of 100 sinusoids where harmonic coefficients are updated at 250 frames per second (FPS) \cite{engel2020ddsp}

B.) \textit{DWTS}: 10 pre-learned wavetables where
wavetable weights are also updated at 250 FPS.

The remaining elements of each approach are assumed identical in specification and performance. There are two clear areas of improvement with our new method. Firstly, \textit{DWTS} requires only ten interpolated wavetable read operations per sample, compared to the 100 required for \textit{Additive}. Secondly, both pipelines require frame-wise linear smoothing of control data to avoid unwanted artefacts in the synthesized signals. \textit{Additive} requires all 100 sinusoidal amplitude coefficients to be smoothed per sample, whereas \textit{DWTS} requires smoothing on only 10. We confirm this on a 2.6GHz 2019 Macbook Pro. Over 10k trials, \textit{Additive} takes on average 251.32ms to generate all 250 frames in 1 second of audio, whereas \textit{DWTS} takes only 20.19ms. 

\section{Conclusion}
\label{sec:conclusion}

In this paper, we presented a differentiable wavetable synthesizer capable of high fidelity neural audio synthesis. We demonstrate how explicit and learnable wavetables offer many advantages including robust one-shot learning and an order of magnitude reduction in computational complexity. The approach opens up applications such as data-efficient neural audio sampling and pitch-shifting.

% -------------------------------------------------------------------------
\bibliographystyle{IEEE}
\bibliography{ICASSP.bib}

\end{document}